\documentstyle[12pt,a4wide]{article}
\newcommand{\be}{\begin{equation}}
\newcommand{\ee}{\end{equation}}
\newcommand{\bea}{\begin{eqnarray}}
\newcommand{\eea}{\end{eqnarray}}
\newcommand{\bean}{\begin{eqnarray*}}

\newcommand{\eean}{\end{eqnarray*}}
\font\upright=cmu10 scaled\magstep1
\font\sans=cmss10
\newcommand{\ssf}{\sans}
\newcommand{\stroke}{\vrule height8pt width0.4pt depth-0.1pt}
\newcommand{\Z}{\hbox{\upright\rlap{\ssf Z}\kern 2.7pt {\ssf Z}}}

\newcommand{\C}{{\rlap{\rlap{C}\kern 3.8pt\stroke}\phantom{C}}}
\newcommand{\R}{\hbox{\upright\rlap{I}\kern 1.7pt R}}
\newcommand{\CP}{\C{\upright\rlap{I}\kern 1.5pt P}}
\newcommand{\PP}{\hbox{\upright\rlap{I}\kern 1.5pt P}}

\newcommand{\identity}{{\upright\rlap{1}\kern 2.0pt 1}}

\newcommand{\bphi}{\mbox{\boldmath $\phi$}}

\newcommand{\HH}{\mbox{\hbox{\upright\rlap{I}\kern 1.7pt H}}}

\newcommand{\LL}{Landau-Lifshitz}

\newcommand{\news}{\setcounter{equation}{0}}
\input{epsf}
\begin{document}
\title{\vskip -70pt
\begin{flushright}
{\normalsize To appear in Physica D} \\
\end{flushright}\vskip 50pt
{\bf \Large \bf SOLITON DYNAMICS IN 3D FERROMAGNETS}\\[30pt]
\author{Theodora Ioannidou and Paul M. Sutcliffe\\[10pt]
\\{\normalsize  {\sl Institute of Mathematics, University of Kent at Canterbury,}}\\
{\normalsize {\sl Canterbury, CT2 7NF, U.K.}}\\
{\normalsize{\sl Email : T.Ioannidou@ukc.ac.uk}}\\
{\normalsize{\sl Email : P.M.Sutcliffe@ukc.ac.uk}}\\}}
\date{November 2000}
\maketitle

\begin{abstract}
\noindent We study the dynamics of solitons in a Landau-Lifshitz
equation describing the magnetization of a three-dimensional ferromagnet
with an easy axis anisotropy. We numerically compute the energy dispersion
relation and the structure of moving solitons,
 using a constrained minimization algorithm. We compare the results with
those obtained using an approximate form for the moving soliton, valid in the 
small momentum limit. We also study
 the interaction and scattering of two solitons, through a numerical
simulation of the (3+1)-dimensional equations of motion.
We find that the force between two solitons can be either attractive or
repulsive depending on their relative internal phase and that 
in a collision two solitons can form an unstable
 oscillating loop of magnons.

\end{abstract}

\newpage
\section{Introduction}
\news\ \ \ \ \ \
In the continuum approximation the state of a ferromagnet is
described by a three component unit vector, $\bphi({\bf x},t)$,
 which gives the local orientation of the magnetization.
The dynamics of the ferromagnet, in the absence of dissipation,
 is goverened by the \LL\ equation
\be
\frac{\partial \bphi}{\partial t}=-\bphi \times 
\frac{\delta E}{\delta \bphi}
\label{ll}
\ee
where $E$ is the magnetic crystal energy of the ferromagnet.
We have chosen units in which the spin stiffness and
magnetic moment density of the ferromagnet are set to one.

The case we study in this paper is that of a three-dimensional
ferromagnet with isotropic exchange interactions and an easy-axis
anisotropy, in which case the energy is given by
\be
E= \frac{1}{2} \int (\partial_i\bphi\cdot\partial_i\bphi 
+ A (1-\phi_3^2)) \ d^3{\bf x}
\label{energy}
\ee
where $A>0$ is the anisotropy parameter and we choose the ground state
to be $\bphi=(\phi_1,\phi_2,\phi_3)=(0,0,1)={\bf e}_3.$

In this case the \LL\ equation becomes
\be
\frac{\partial \bphi}{\partial t}=\bphi \times 
(\partial_i\partial_i\bphi + A\phi_3{\bf e}_3).
\label{ll3}
\ee

This equation has finite energy, stable, exponentially localized
solutions known as magnetic solitons \cite{KIK}. In section 2
we review the properties of stationary magnetic solitons
and in section 3 we numerically compute solutions describing
moving solitons, display their energy dispersion relation and discuss
their structure.
Finally, in section 4, we perform numerical simulations of the
time dependent equations of motion to investigate the interaction
and scattering of two solitons. We find that the force between
two solitons depends on their relative internal phase, and that
during a collision two solitons can form an unstable  loop of magnons
which subsequently decays into solitons.

\section{Stationary Solitons}
\news\ \ \ \ \ \
\begin{figure}[ht]
\begin{center}
\leavevmode
\epsfxsize=10cm\epsffile{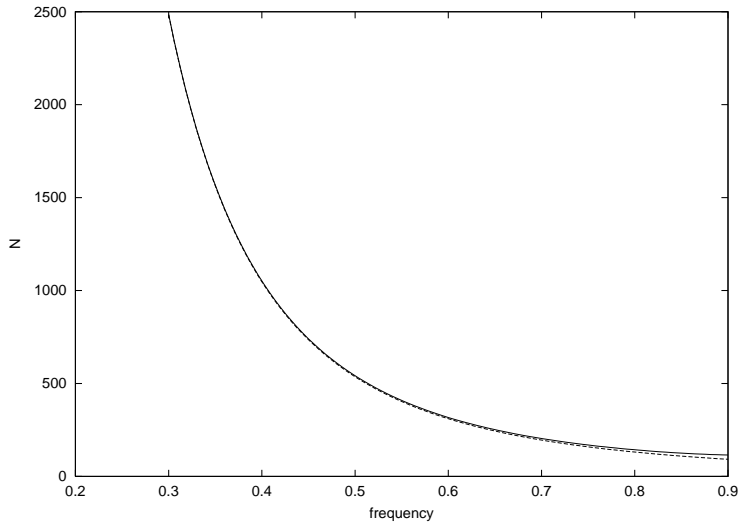}
\caption{The number of spin reversals, $N$, as a function
of the precesion frequency $\omega$. The solid curve is the
numerical result and the dashed curve is the approximate formula (\ref{anform}).}
\label{fig1}
\end{center}
\end{figure}

In addition to the energy (\ref{energy}), the \LL\ equation (\ref{ll3})
has two other conserved quantities. These are the number of spin
reversals, $N$, and the momentum ${\bf P}$,
 given by \cite{Pap}
\be
N=\int (1-\phi_3) \  d^3{\bf x}
\ee
and 
\be
P_i=\frac{1}{4}\epsilon_{ijk}\int x_j \epsilon_{klm}
\bphi\cdot\partial_l\bphi\times\partial_m\bphi \ d^3{\bf x}.
\label{mom}
\ee 
In the quantum description, $N$ counts the number of quasi-particles
in the magnet, that is, it may be interpreted as the magnon number.

In this section we consider only stationary solitons, ie
${\bf P}=0$, whose properties have been discussed in ref.\cite{KIK}.
There are no static solitons, but stationary solitons have a time
dependence which resides only in the constant motion of an internal phase.
Explicitly, the stationary soliton has the form
\be
\bphi=\frac{1}{1+f^2}(2f\cos(\omega t),-2f\sin(\omega t),1-f^2)
\ee
where $\omega$ is the frequency and $f(r)$ is a real profile function 
with the boundary conditions
$f(\infty)=0$, $f'(0)=0$. The resulting
equation for the profile function is
\be
f''=-\frac{2f'}{r}+\frac{2ff'^2+Af(1-f^2)}{1+f^2}-\omega f
\label{pro}
\ee
For large $r$ this equation may be linearized to give the asymptotic
solution
\be
f\sim B \exp(-r\sqrt{A-\omega})
\ee
from which it can be seen that a soliton solution exists only for
$\omega<A.$ It can also be shown that $\omega>0$ and hence an anisotropy
term is vital for the existence of these stationary solitons.
From now on we set $A=1$ so that $0<\omega<1.$
There is a one-parameter family of stationary solitons parameterized
by either the frequency $\omega$ or the number of spin reversals
$N.$ In figure 1 we plot the relation between these
 two quantities,
obtained by solving equation (\ref{pro}) using a shooting method (solid curve),
and in figure 2 we display (solid curve) the energy per spin 
reversal of the soliton,
 $E/N$ as a function of $N.$ 

\begin{figure}[ht]
\begin{center}
\leavevmode
\epsfxsize=10cm\epsffile{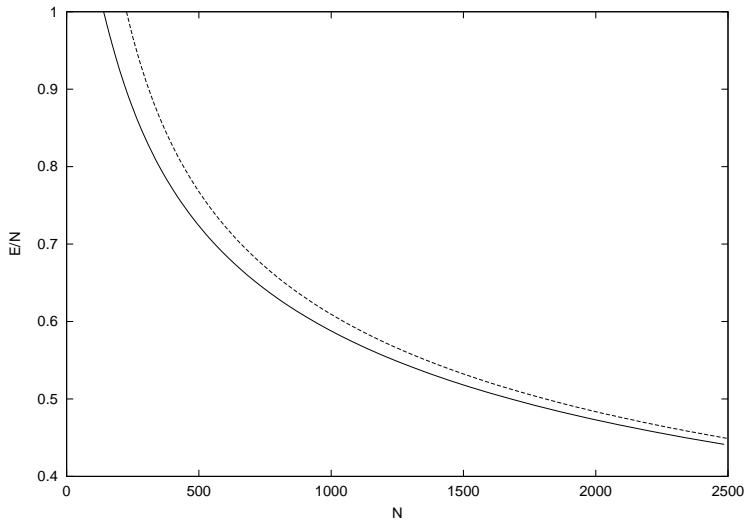}
\caption{The energy per spin reversal, $E/N$, as a function
of the spin reversal $N$. The solid curve is the
numerical result and the dashed curve is the approximate formula (\ref{anform}).}
\label{fig2}
\end{center}
\end{figure}

In the limit of small $\omega$ (which corresponds to large $N$)
the solution is approximately given by $f=e^{R-r}$, where $R\gg 1$ is
a core radius which marks the boundary between the two vacua
$\bphi=\pm{\bf e}_3$ inside and outside the soliton. In this
case the soliton may be thought of as a spherical domain wall. 
Using this simple approximate solution the following
asymptotic formula may be derived, which are valid for large $N$ 
\be
N=\frac{64\pi}{3\omega^3}, \ \ 
E=\frac{32\pi}{\omega^2}=(72\pi N^2)^{1/3}.
\label{anform}
\ee
The dashed curves in figures 1 and 2 are obtained from these asymptotic
formulae and fit the numerical results well.

\section{Moving Solitons}
\news\ \ \ \ \ \
To compute moving solitons is a more difficult task than in the
stationary case and can not be reduced to simply solving an ordinary
differential equation. There is an ansatz which is consistent with
the equations of motion and describes a soliton which moves at
constant velocity ${\bf v}$ and rotates in internal space with
frequency $\omega.$ Explicitly the ansatz reads
\be
\phi_1+i\phi_2=(\widehat\phi_1({\bf x}-{\bf v}t)
+i\widehat\phi_2({\bf x}-{\bf v}t))e^{-i\omega t}, \ \
\phi_3=\hat\phi_3({\bf x}-{\bf v}t).
\label{trav}
\ee
Substituting this ansatz into the equation of motion
(\ref{ll3}) leads to a partial differential equation for
$\mbox{\boldmath $\hat\phi$}$
 which is not compatable with a spherically symmetric
solution for non-zero ${\bf v}.$ This partial differential equation
has a variational formulation, which is useful in computing its
solutions. Let $\hat E({\bf P},N)$ be the minimal value of $E$ for fixed
values of the momentum ${\bf P}$ and number of spin reversals $N$.
Then the solution of this constrained minimization problem is
precisely the function 
$\mbox{\boldmath $\hat\phi$}$
corresponding to the values
\cite{TW}
\be
\omega=\frac{\partial\hat E}{\partial N}\bigg\vert_{\bf P}, \ \
v_i=\frac{\partial\hat E}{\partial P_i}\bigg\vert_N.
\label{speed}
\ee

We compute moving soliton solutions by numerically solving
this constrained minimization problem. A similar computation has been
performed in the case of an isotropic magnet ($A=0$) \cite{Coo}, but the results
in this case are qualitatively different. For example, we have already
seen that without anisotropy there are no stationary solutions, and in fact
moving solitons exist in the isotropic case only for sufficiently 
large momenta \cite{Coo}.

A moving soliton has an axial symmetry in the plane perpendicular
to its momentum ${\bf P}.$ Since the \LL\ equation is $SO(3)$
invariant we can, without loss of generality, choose the momentum to be
in the $x_3$ direction, ${\bf P}=(0,0,P),$ so that the soliton
moves along the $x_3$ axis and has an axial symmetry in the $x_1x_2$ plane.
We use cylindrical coordinates, with $\rho=\sqrt{x_1^2+x_2^2}$, and discretize
the energy using second order differences and a grid of size
$50\times 100$ in the $\rho,x_3$ plane. The energy is minimized using
a simulated annealing algorithm \cite{sabook,HSW} and the constraints
on $N$ and $P$ are imposed using a penalty function method.

\begin{figure}[ht]
\begin{center}
\leavevmode
\epsfxsize=10cm\epsffile{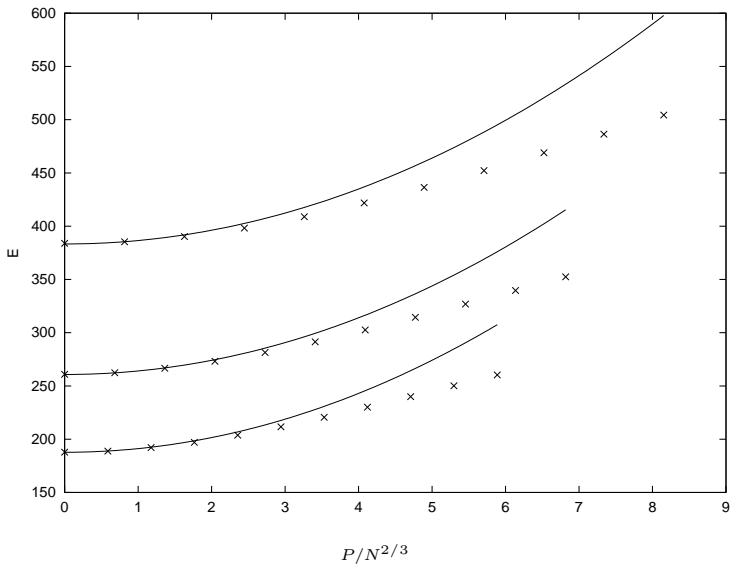}
\ \vskip 0cm {\bf\tiny\bf $P/N^{2/3}$} \vskip 0cm 
\caption{
The energy, $E$, as a function of the scaled momentum,
$P/N^{2/3}$, for the three values $N=204,317,542.$ The
crosses represent
the numerical data and the curves are obtained from the small momentum
approximation.}
\label{fig3}
\end{center}
\end{figure}

\begin{figure}[ht]
\begin{center}
\leavevmode
\epsfxsize=10cm\epsffile{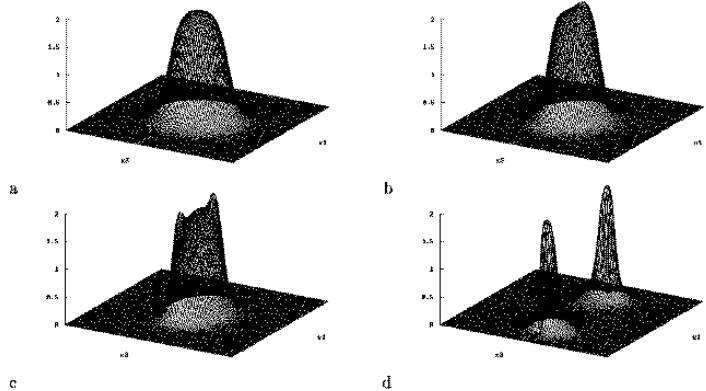}
\ \vskip 0cm
\caption{
The spin reversal density, $1-\phi_3$, in the $x_2=0$ plane,
for the soliton with $N=317$, and four values of the scaled momentum,
$P/N^{2/3}=0,6.8,13.6,68.2$}
\label{fig4}
\end{center}
\end{figure}

In figure 3 we plot (crosses) the energy dispersion relation, $\hat E(P,N)$
obtained from these computations for the three values $N=204,317,542.$
The energy increases with $N$ so the data need not be labelled individually.
Note that the rather unobvious choice for these three values is due to the
 fact that in the zero momentum case they correspond to the frequencies 
$\omega=0.7,0.6,0.5.$ The soliton velocity, if required, can be read-off
from this data by making use of formula (\ref{speed}).

We now describe an approximate initial condition for the field
$\bphi$ to describe a soliton with any given values of $N$ and $P$ 
and assess its accuracy by computing its energy dispersion relation
and comparing with the results displayed in figure 3. 

It was noted in ref.\cite{PZ} that in the two-dimensional case
a moving soliton could be generated from initial conditions which
consist of performing a space dependent phase transformation
on the stationary soliton solution, although no relation was given
between the parameter which appears in this transformation and 
the momentum of the resulting soliton. In the following we 
study the same type of transformation in the three-dimensional case, 
although we have a simple explicit form for the parameters $N,P$ of
the corresponding solution and this will allow us to compute the 
energy dispersion relation to compare with the previous numerical results.

Our approximate initial conditions are given by
\be
\bphi=\frac{1}{1+f^2}(2f\cos(Px_3/N),2f\sin(Px_3/N),1-f^2)
\label{ansatz}
\ee
where $f(r)$ is the stationary soliton profile function 
with $N$ spin reversals, which we have already introduced earlier. 
Note that since $\phi_3$ is the same as in the stationary case then
the number of spin reversals of this field is indeed $N.$
We now show that the momentum of this field is ${\bf P}=(0,0,P).$
By symmetry it is trivial to show that $P_1=P_2=0.$ Substituting
(\ref{ansatz}) into the momentum equation (\ref{mom}), using spherical
variables and performing the angular part of the integration we have
\be
P_3=-\frac{16\pi P}{3N}\int_0^\infty \frac{ff'r^3}{(1+f^2)^2} \  dr
=\frac{8\pi P}{N}\int_0^\infty \frac{f^2r^2}{(1+f^2)}
-\frac{1}{3}\big(\frac{f^2r^3}{(1+f^2)}\big)' \ dr
=P
\ee 
where we have used the boundary conditions on $f(r)$ to set the total
derivative term to zero and recognized in the remaining term the
expression for $N.$

Having verified that this configuration has $N$ spin reversals and
momentum $P$ we can now compute its energy and derive its dispersion
relation. Substituting the field (\ref{ansatz}) into the expression
for the energy (\ref{energy}) and performing the angular integrals
we find
\be
E=E_0(N)+P^2E_1(N), 
\label{approxe}
\ee
where
\be
E_0(N)=8\pi\int_0^{\infty} \frac{(f'^2+f^2)r^2}{(1+f^2)^2} \  dr ,
\ \hskip 1cm
E_1(N)=\frac{8\pi}{N^2}\int_0^{\infty} \frac{f^2r^2}{(1+f^2)^2} \  dr
\ee
The dispersion relation obtained from equation (\ref{approxe})
is also displayed in figure 3 (solid curves) for comparison with the
full numerical results. 
In ref.\cite{Coo} it is noted that a magnon loop occurs for
$P\gg N^{2/3}$, and from figure 3 
it can be seen that the ansatz is only a good approximation if the momentum
 stays below this range, since otherwise it
begins to lose its accuracy and has considerably more energy
than the true solution.

In figure 4 we plot, in the $x_2=0$ plane,
 the spin reversal density, $1-\phi_3$, for the soliton with
$N=317$ and four different values of the scaled momentum,
$P/N^{2/3}=0,6.8,13.6,68.2$

Recall that in the spherical ansatz (\ref{ansatz}) the spin reversal
density is assumed to be the same as in the stationary case. Therefore by
examining how the spin reversal density changes with increasing momentum
we can assess the validity of the ansatz. 
By a comparison of figures 4a and 4b, it can be seen that the spherical assumption
in the ansatz is substantially violated for $P\sim N^{2/3}$, with the true
soliton being stretched in the plane orthogonal to its motion.
For very large momenta, $P\gg N^{2/3}$, there is a transition from a single lump
of magnons into a magnon loop. Already at $P/N^{2/3}=13.6$, figure 4c, one can see that
the soliton is sufficiently distorted so that the maximum
of the spin reversal density is no longer on the $x_3$ axis. For $P/N^{2/3}=68.2$,
figure 4d, the soliton has now clearly formed two distinct components, which
when we recall the axial symmetry in the $x_1x_2$ plane, means that the 
spin reversal density is now localized around a circle in a plane perpendicular
to its motion. In the isotropic case all the solitons have this
structure \cite{Coo} and are known as magnetic vortex rings.

In summary, we have seen that low momentum solitons can be described by a
spherical ansatz but as the momentum increases there is a transition from
lump-like solitons to ring-like solitons.

\section{Multi-Soliton Interactions}
\news\ \ \ \ \ \
In this section we discuss the results of a numerical evolution of
the full time-dependent \LL\ equation (\ref{ll3}) in order to
investigate the interaction and scattering of two solitons.

Two initially stationary and well-separated solitons have an axial
symmetry about the line connecting them, and we make use of this
symmetry in our numerical evolution code. If we take two solitons
 on the $x_3$-axis, each with momentum only in the $x_3$ direction,
 then both the initial conditions and the equations of motion have an axial
symmetry in the $x_1x_2$ plane. We could use cylindrical coordinates
and evolve the equations of motion in these variables but this type
of implementation can often lead to numerical instabilities associated
with coordinate singularities, particularly along the $x_3$-axis.
We therefore employ the \lq Cartoon\rq\ method recently proposed in
ref.\cite{cartoon} in the context of solving the axisymmetric 
Einstein equations. In this approach the equations are evolved in 
cartesian coordinates, but on a thin slab, that is the grid
has size $M\times 3\times M$ so that only 3 points are used in the $x_2$
direction. In the centre of the slab, which is the plane $x_2=0$, the
equations of motion are evolved in a standard manner; we employ a 
fourth-order Runge-Kutta method for the time evolution and space
derivatives are approximated by second order finite differences.
Away from this central slice the field values are determined
by making use of the axial symmetry to map to an equivalent point
in the central slice, $x_2=0.$ Since the field values in this
central slice are only known on a discrete lattice a one-dimensional
 interpolation algorithm must be employed, which we take to be a simple
second order Lagrange interpolant. The results in this section were
obtained using grids of size $M=201$, with time and space steps given
by $\Delta t=0.025, \Delta x=0.3.$

To obtain an initial condition consisting of two well-separated
solitons we use the following nonlinear superposition rule.
Given the $S^2$-valued field $\bphi$ corresponding to the
first soliton we construct the associated Riemann sphere variable
obtained by stereographic projection of the $S^2$-valued field
onto the complex plane. We construct a similar Riemann sphere field
for the second soliton and obtain the combined $S^2$ field by addition
of the two Riemann sphere fields followed by an inverse stereographic
 projection. Note that before the two fields are combinded a relative phase,
$\alpha$, can be introduced corresponding to rotating the $\phi_1,\phi_2$
components of the first soliton through an angle $\alpha.$ As we shall see,
this relative phase has a marked influence on the force between the two
solitons.

As an initial condition we take two solitons, each corresponding
to a stationary soliton with $N=317$, placed on the $x_3$-axis at 
the positions $x_3=\pm 7$. Figure 5 diplays the subsequent evolution
of their relative separation for the cases when there is no relative
phase, that is $\alpha=0$, (bottom curve) and when the two solitons
are exactly out of phase, that is $\alpha=\pi$, (top curve).
In computing the separation of the two solitons the position of each
soliton is defined as the point at which the spin-reversal density,
 $1-\phi_3$, takes its maximum value, and this is calculated using a
quadratic interpolation around the lattice site at which it takes
its maximal value.

\begin{figure}[ht]
\begin{center}
\leavevmode
\epsfxsize=10cm\epsffile{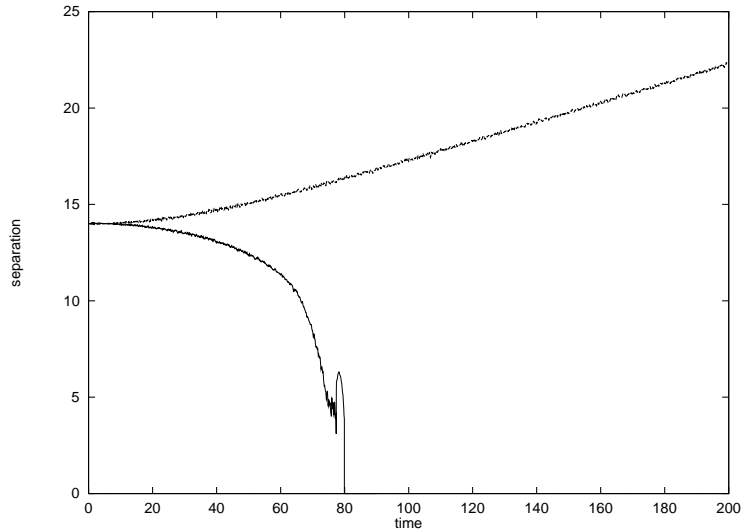}
\ \vskip 0cm
\caption{The time evolution of the separation between two
initially stationary solitons. The relative internal phase, $\alpha$,
is set to 0 (bottom curve) and $\pi$ (top curve).}
\label{fig5}
\end{center}
\end{figure}

From figure 5 we see that when there is no relative phase the two
solitons attract and form a single larger soliton, which is perhaps 
what is expected from the fact that the energy per spin reversal is
a decreasing function of $N$ as shown in figure 2. The slight peak
before the solitons merge in figure 5 is a result of the highly 
nonlinear deformation the solitons suffer as they combine.
In contrast, when the solitons are out of phase they repel
and quickly settle to a state in which the two solitons are moving away
from each other at a constant speed. Thus we have demonstrated that
the force between two solitons can be attractive or repulsive
depending on their relative phase. 

If the relative phase is something
other than $0$ or $\pi$ then the situation is even more complicated.
In this case the ultimate fate of the solitons, that is whether the final
configuration consists of one or two solitons, depends on the relative phase
and initial separation of the solitons. However, the picture is not as simple
as a mere attraction or repulsion, as the solitons can initially move towards
each other but then turn around and ultimately repel. Furthermore, during this
interaction there can be a magnon exchange, so that the number of spin reversals
of one soliton can increase while that of the other decreases. Qualitatively
similar results have been found and studied in great detail \cite{BS} for solitons
known as Q-balls, where the role of the number of spin reversals is played
by a Noether charge. Q-balls occur in relativistic field theories, in which the dynamics
is second order in time, rather than the first order dynamics of the \LL\ 
equation, so it may seem a little surprising that there should be similarities.
However, Q-balls have a time-dependent internal phase, very similar to the
precession of the magnetization for the magnetic solitons studied in this paper,
and as argued in ref.\cite{BS}, to which we direct the reader for further
details, the novel dynamics can be mainly attributed to this fact.
As we discuss below, there are other common features between the dynamics
of magnetic solitons and Q-balls.

For two solitons which individually have no momentum and are in-phase
we have seen that the two solitons merge and, after some oscillations,
settle down to a single soliton which has a value of $N$ which is
approximately the sum of its constituents.
A small amount of energy is radiated during these oscillations, in the form of magnons
which are released to infinity, and can be dealt with numerically
 by applying absorbing boundary conditions at the edge of the grid.
As we now demonstrate, if the solitons initially
have some momentum then a much more complicated evolution takes place.

\ 

\begin{figure}[ht]
\begin{center}
\leavevmode
\epsfxsize=12cm\epsffile{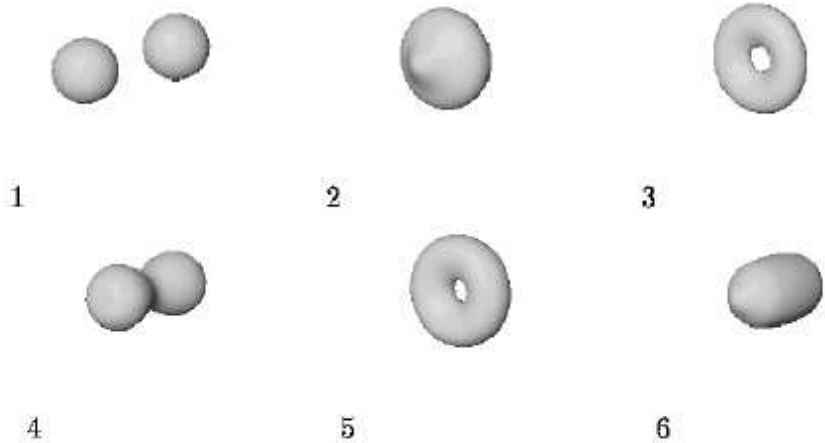}
\ \vskip -0cm
\caption{Isosurface plots of $1-\phi_3=1.5$ at times
$t=0,20,30,65,80,115$ during the collision of two solitons. }
\label{fig6}
\end{center}
\end{figure}

In figure 6 we display a fully three-dimensional isosurface plot corresponding to 
the surface where $1-\phi_3=1.5$, at six different times $t=0,20,30,65,80,115.$
 The initial
conditions were created from two $N=317$ solitons, with zero relative
phase, placed at the positions $x_3=\pm 7$ and with momentum
$P=\mp 158.$ The individual moving solitons were constructed using
the ansatz (\ref{ansatz}), which is an adequate approximation at this
momentum. We see from figure 6 that the two solitons collide and 
form a single lump (figure 6.2) which then expands to 
form an axially symmetric loop of magnons, that is, a toroidal region
where the spin-reversal density is concentrated (figure 6.3). This magnon loop
initially expands until it reaches a critical radius, which increases
with the initial momentum of the solitons, and then contracts to
produce two solitons which move apart along the initial line of
approach (figure 6.4). The solitons then attract once more, reforming the loop
(figure 6.5), which again collapses, with the solitons again attempting
to separate (figure 6.6). This process continues through several cycles,
with the loop forming at a slightly smaller radius each time, until eventually
the configuration settles to a single large soliton with no momentum.
Note that the loop which forms in this process is not the same kind of 
structure as the moving vortex ring, since it has no momentum perpendicular
to the plane of the loop. If the initial momentum of the solitons is
increased slightly (for example, with $P=317$, and all other initial
conditions the same) then the first loop formed is of sufficient radius that
its collapse results in the two solitons which emerge from its decay having
sufficient momentum that they never recombine and instead travel out to infinity.

This phenomenon is the three-dimensional realization of 
the two-dimensional process described in ref.\cite{PZ} where two
solitons scatter at right angles to the initial line of approach.
In the three-dimensional case the axial symmetry prevents the two
solitons from scattering at right angles and results instead in the
formation of a magnon loop. 

This kind of loop formation also appears in relativistic Q-ball dynamics
and the qualitative features are very similar \cite{BS}.

In summary, we have seen that the interaction of multi-solitons,
even at relatively low momenta, is a very complicated process 
with forces which depend on relative internal phases, and novel
features such as the production of unstable loops. Clearly these
processes require further study and hopefully a more analytical
understanding will emerge.

\section{Conclusion}
\news\ \ \ \ \ \
We have used several different numerical techniques to study the 
dynamics and interaction of magnetic solitons in a three-dimensional
ferromagnet with easy-axis anisotropy. We have computed moving solitons
using a minimization algorithm and compared the results to those of 
a simple radial ansatz, which we have shown is a good approximation for
low momenta. However, for large momenta there is a transition from
lump-like solitons to ring-like solitons, where obviously the
radial ansatz fails badly. We have found that the interaction between two solitons
has a strong dependence on their relative phase and that the collision
of solitons can be highly non-trivial and lead to the formation of
unstable magnon loops. Finally, we have observed that many of the
features found for magnetic solitons are qualitatively very similar
to those of Q-balls, despite the fact that the dynamics of the
latter is a second order in time relativistic system.

\section*{Acknowledgements}
\news\ \ \ \ \ \
We acknowledge the Nuffield Foundation for an award (TI) and
 the EPSRC for an Advanced Fellowship (PMS).\\


\end{document}